\documentclass[twocolumn,showpacs,floats,floatfix,aps,prl]{revtex4}
\usepackage{amsfonts,amssymb,amsmath}
\usepackage{color,calc}
\usepackage[dvips]{graphicx}
\usepackage{bm}
\usepackage[normalem]{ulem}

\def\be{ \begin{equation} }
\def\ee{ \end{equation} }
\def\bea{ \begin{eqnarray} }
\def\eea{ \end{eqnarray} }
\def\bse{ \begin{subequations} }
\def\ese{ \end{subequations} }

\def\i{\,\text{i}}
\def\e{\,\text{e}}
\def\i{i}
\def\e{e}

\def\to{\rightarrow}
\def\fromto{\leftrightarrow}

\newcommand{\ket}[1]{\vert #1\rangle}
\def\d{\text{d}}

\def\U{\mathbf{U}}

\def\H{\mathbf{H}}
\def\c{\mathbf{c}}

\def\C{C} 
\def\S{S} 

\def\black{}

\def\rms{\textsc{rms}}

\def\sec{\textbf}

\def\ket#1{| #1 \rangle}
\def\bra#1{\langle #1 |}
\def\proj{\Pi}
\def\phase{\phi}
\def\phaseV{\phase}

\def\from{\leftarrow}
\def\to{\rightarrow}

\begin{document}

\author{Genko T. Genov}
\affiliation{Department of Physics, Sofia University, 5 James Bourchier blvd, 1164 Sofia, Bulgaria}
\author{Nikolay V. Vitanov}
\affiliation{Department of Physics, Sofia University, 5 James Bourchier blvd, 1164 Sofia, Bulgaria}
\title{Dynamical suppression of unwanted transition paths in multistate quantum systems}
\date{\today}

\begin{abstract}
We introduce a method to suppress unwanted transition channels, even without knowing their couplings,
  and achieve perfect population transfer in multistate quantum systems by the application of composite pulse sequences.
Unwanted transition paths may be present due to imperfect light polarization, stray electromagnetic fields,
 misalignment of quantization axis, spatial inhomogeneity of trapping fields, off-resonant couplings, etc.
Compensation of simultaneous deviations in polarization, pulse area, and detuning is demonstrated.
The accuracy, the flexibility and the robustness of this technique make it suitable for high-fidelity applications in quantum optics and quantum information processing.
\end{abstract}

\maketitle

\sec{Introduction.}
%
Experiments in various branches of quantum physics require well-defined quantum energy states and well-defined interactions.
For example, one of the basic conditions for the future quantum computer is a well defined qubit --- a two-state quantum system.
Real quantum systems, however, usually possess a huge number of quantum states and special care is needed to isolate just two of them.
In real and artificial atoms this is usually done with polarized laser light, carefully aligned with the quantization axis.
However, unwanted transition channels may still be present, which reduce the fidelity of the operations.
For example, when an ultracold atomic ensemble held in a dipole trap is addressed by right circularly polarized ($\sigma^+$) light, it is difficult to have all atoms ``seeing'' the same $\sigma^+$ polarization
 because not all of them are exactly in the focus of the laser fields; consequently, many atoms will ``see'' an admixture of $\sigma^+$ and $\sigma^-$ light (i.e., elliptically polarized light),
 which limits the fidelity of quantum state control.
Unwanted transition paths may be present also due to imperfect polarization or alignment, stray electric and/or magnetic fields, off-resonant couplings, etc.

\begin{figure}[t]
\begin{center}
 \includegraphics[width=0.85 \columnwidth]{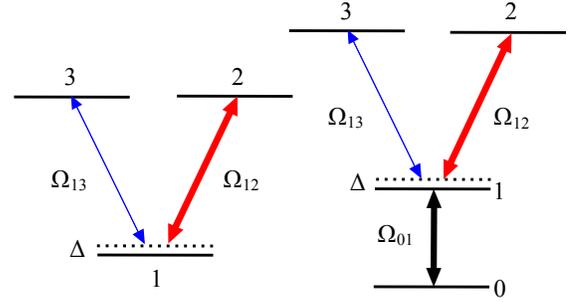}
 \end{center}
 \caption{(color online) V system (left) and Y system (right).
  }
\label{Fig-VY}
\end{figure}

In this paper, we propose a simple and efficient technique for automatic compensation of such errors, which uses composite pulse sequences to dynamically suppress unwanted transition channels even without knowing the magnitude of these errors.
We illustrate the technique in three-state and four-state quantum systems forming linkages reminiscent of the letters ``V'' and ``Y'', as shown in Fig. \ref{Fig-VY};
 the technique is, however, applicable also to more complex linkage patterns.
We demonstrate compensation of independent as well as simultaneous variations in polarization, pulse area and detuning.

Composite pulse sequences have been used for several decades in nuclear magnetic resonance \cite{Levitt86,Freeman97}, and since recently, in quantum information processing \cite{Haffner,RoosMolmer,Hill,Ivanov11PRA} and quantum optics \cite{Torosov11PRA,Ivanov11OL,Torosov11PRL,Genov11PRA} as a versatile control tool for quantum systems.
In fact, some of the basic ideas have been developed earlier in polarization optics in research on achromatic polarization retarders \cite{Optics,WolfAzzamGoldstein}.
While composite pulse sequences have been used mainly for two-state quantum systems, there are also studies of three-state and multistate systems \cite{Steffen03,Genov11PRA}.

A composite pulse is a sequence of pulses with well defined relative phases, 
 which are used as control parameters in order to compensate imperfections in the excitation profile produced by a single pulse, or to shape the profile in a desired manner.
The imperfections may be caused by an imprecise pulse area (due to fluctuating field intensity and pulse duration, spatial field inhomogeneity, etc.), undesirable frequency offset (due to uncompensated electric and magnetic fields, dynamic Stark shifts, Doppler shifts, etc.), or unwanted frequency chirp.

Here we use the toolbox of composite pulses to design recipes for compensation of imperfect light polarization, which may open unwanted transition paths and may turn a qubit or a simple three-state ladder into a complex tree of multiple states, with an unavoidable loss of efficiency.

\sec{V system.} \label{SecV}
The dynamics of a coherently driven V-shaped quantum system, shown in Fig.~\ref{Fig-VY} (left), obeys the Schr\"{o}dinger equation,
\be\label{Schrodinger equation}
\i \hbar\partial_t \mathbf{c}(t) = \H(t)\mathbf{c}(t),
\ee
where the vector $\c(t) = [c_1(t), c_2(t), c_3(t)]^T$ contains the probability amplitudes of the three states.
The Hamiltonian in the rotating-wave approximation (RWA) reads
\begin{align}
\mathbf{H}_V(t) &= (\hbar/2) \Delta (\proj_{11} - \proj_{22} - \proj_{33})  \notag\\
 & + (\hbar/2)\left[\Omega_{12}(t)\e^{\i\phaseV_{12}} \proj_{12} + \Omega_{13}(t)\e^{\i\phaseV_{13}} \proj_{13} +\text{h.c.} \right],
  \label{HamiltonianV1}
\end{align}
where $\Delta=\omega_0 -\omega$ is the detuning between the laser carrier frequency $\omega$ and the Bohr transition frequency $\omega_0$, and $\proj_{jk} = \ket{j}\bra{k}$.
The magnitudes of the Rabi frequencies are $\Omega_{jk}(t)=|\mathbf{d}_{jk}\cdot \mathbf{E}(t)|/\hbar$,
 where $\mathbf{E}(t)$ is the envelope of the laser electric field
 and $\mathbf{d}_{jk}$ is the transition dipole moment of the respective transition $j\leftrightarrow k$;
 the phases of the Rabi frequencies are $\phaseV_{12}$ and $\phaseV_{13}$.
We assume that the two Rabi frequencies have the same time dependence $f(t)$ and we introduce the root-mean-square (\rms) peak Rabi frequency $\Omega$ and the mixing angle $\theta$ via $\Omega_{12}(t) = \Omega f(t) \cos\theta$ and  $\Omega_{13}(t) = \Omega f(t) \sin\theta$.

An important example of such a linkage is the transition between the magnetic sublevel $m=0$ (state $\ket{1}$) of a ground level with an angular momentum $j=0$ or 1, and the magnetic sublevels $m=1$ (state $\ket{2}$) and $m=-1$ (state $\ket{3}$) of an excited level with an angular momentum $j=1$ driven by an elliptically polarized laser pulse.
The latter can be represented as a superposition of two circularly polarized $\sigma^+$ and $\sigma^-$ pulses \cite{WolfKlein}:
 then, $\Omega_{12}=\Omega_+$ and $\Omega_{13}=\Omega_-$, the angle of rotation of the polarization ellipse is $\phaseV=(\phaseV_{12} - \phaseV_{13})/2$ and the ellipticity is $\varepsilon = \cos 2\theta$.
 The values $\theta=0,\pi/4,\pi/2$ ($\varepsilon = 1,0,-1$) correspond, respectively, to $\sigma^+$, linear and $\sigma^{-}$ polarizations.

Our objective is to transfer all population from state $\ket{1}$ to state $\ket{2}$ and completely suppress the excitation channel $\ket{1} \leftrightarrow \ket{3}$.
In the above example of magnetic sublevels, this can be achieved by a $\sigma^+$ polarized $\pi$ pulse.
However, if the polarization is not perfectly $\sigma^+$, then the atom will ``see'' some $\sigma^-$ polarized light and the unwanted excitation channel $\ket{1} \leftrightarrow \ket{3}$ will open up.
We show below that composite pulse sequences, instead of a single $\pi$ pulse, can compensate such an admixture of unwanted polarization,
 even without knowing its amount, and achieve perfect transfer efficiency.

The V system described by the Hamiltonian \eqref{HamiltonianV1} can be transformed by the Morris-Shore transformation \cite{MorrisShore} into a decoupled state $\ket{d} = -\e^{- \i\phaseV_{13}}\sin\theta\ket{2} + \e^{-\i\phaseV_{12}}\cos\theta\ket{3}$
 and a two-state system composed of state $\ket{1}$ and a coupled state $\ket{c} = \e^{\i\phaseV_{12}}\cos\theta\ket{2} + \e^{\i\phaseV_{13}}\sin\theta\ket{3}$ driven by the following Hamiltonian
\be\label{HamiltonianVMS}
\mathbf{\widetilde{H}}_2(t) = (\hbar/2) \{ \Delta (\proj_{11}-\proj_{cc}) + [\Omega f(t) \proj_{1c} + \text{h.c.}]\}.
\ee
The corresponding propagator can be expressed in terms of the complex Cayley-Klein parameters $a$ and $b$ (with $|a|^2+|b|^2=1$) as \cite{Vitanov98}
\be \label{PropagatorVMS}
\mathbf{\widetilde{U}} = \left[\begin{array}{cc} a & b \\ -b^{\star} & a^{\star} \end{array}\right].
\ee
For resonant pulses ($\Delta=0$), with \rms~area $A=\int_{t_i}^{t_f} \Omega f(t) \d t$, the Cayley-Klein parameters are independent of the pulse shape: $a=\cos{A/2}$ and $b=-\i\sin{A/2}$.
For $\Delta\neq 0$, $a$ and $b$ depend on the pulse shape.

The propagator in the original basis reads \cite{Vitanov98,KyosevaVitanov,Genov11PRA}
\begin{align}
& \mathbf{U}(\bm{\phase}) =\label{U_v} \notag\\
& \left[ \begin{array}{ccc}
a & b\e^{\i\phaseV_{12}}\C & b\e^{\i\phaseV_{13}}\S \\
-b^{\star}\,\e^{-\i\phaseV_{12}}\C & a^{\star}\C^2 + \zeta\S^2 & (a^{\star}-\zeta)\,\e^{-2\i\phaseV}\S\C \\
-b^{\star}\,\e^{-\i\phaseV_{13}}\S & (a^{\star}-\zeta)\,\e^{2\i\phaseV}\S\C & \zeta\C^2 + a^{\star}\S^2
\end{array} \right],
\end{align}
where $\S=\sin\theta$, $\C=\cos\theta$
 and $\zeta=\exp[\i\int_{t_i}^{t_f} \Delta(t) \d t/2]$.
Complete population transfer $\ket{1}\to\ket{2}$ with a single pulse implies $|U_{21}|=1$, which therefore requires $a=0$, $|b|=1$ and $\theta=0$.
However, if $\theta\ne 0$, then the coupling between states $\ket{1}$ and $\ket{3}$ is nonzero and some population is unavoidably lost from state $\ket{2}$:
 either transferred to state $\ket{3}$ or left in state $\ket{1}$, because it is impossible to have $|U_{21}|=1$ for $\theta\neq 0$ (recall that $|b|\leqq 1$).

\begin{figure}[t]
\begin{center}
 \includegraphics[width=0.9 \columnwidth]{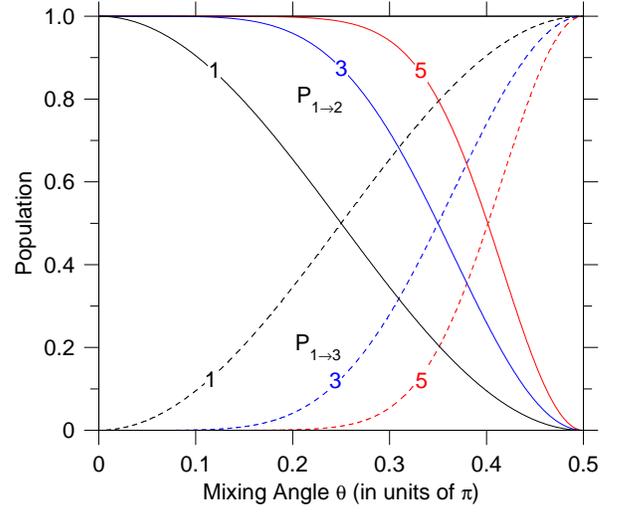}
 \end{center}
 \caption{(color online) Transition probabilities $P_{1\to 2}$ (solid curves) and $P_{1\to 3}$ (dashed curves) for a V system
  vs the mixing angle $\theta$ for a single pulse with \rms~area $\pi$,
  and for composite sequences of three pulses (each with \rms~area $\pi$) with phases
    $\mathbf{\phi}_{12}=(0, 2/3, 0)\pi$; $\mathbf{\phi}_{13}=(0, 1, 1/3)\pi$,
    and five pulses with phases $\mathbf{\phi}_{12}=(0, 1.411, 0.249, -0.432, -0.935)\pi$ and $\mathbf{\phi}_{13}=(0, 0.454, -0.632, 0.14, -0.514)\pi$.
  }
\label{Fig-V}
\end{figure}

Deviation of $\theta$ from 0 can be compensated to an arbitrary order by composite pulses.
The propagator of a composite sequence of $n$ pulses reads
\be
\U^{(n)} = \U(\bm{\phi}_{n})\cdots\U(\bm{\phi}_{2})\U(\bm{\phi}_{1}), \label{U_Comp}
\ee
where $\bm{\phi}_{k}=(\phi_{12}^{(k)},\phi_{13}^{(k)})$ are phase shifts of the $k$-th pulse in the sequence with $\U(\bm{\phi}_{k})$ given by Eq.~\eqref{U_v}. The phases $\phi_{1j}$ \black are free parameters.
We choose to determine them by setting $P_{1\to 2}=|U^{(n)}_{21}|^2=1$ for $\theta=0$ and nullifying the coefficients in the Taylor expansion of $P_{1\to 2}$ vs $\theta$ to the highest possible order.
Since the global phase is irrelevant, we take $\mathbf{\phi}_1=\mathbf{0}$ without loss of generality.
This compensation is demonstrated in Fig.~\ref{Fig-V} for composite sequences of resonant pulses. For longer sequences (larger $n$), the transition profile $P_{1\to 2} (\theta)$ widens and the unwanted transition path  $\ket{1}\to\ket{3}$ is suppressed for a wider range of $\theta$.
Remarkably, for sufficiently long composite sequences, the transition $\ket{1}\to\ket{3}$ can be suppressed even if its coupling is larger than that of the transition $\ket{1}\to\ket{2}$, i.e. in the range $\theta>\pi/4$.

\begin{figure}[t]
\begin{center}
 \includegraphics[width=0.95\columnwidth]{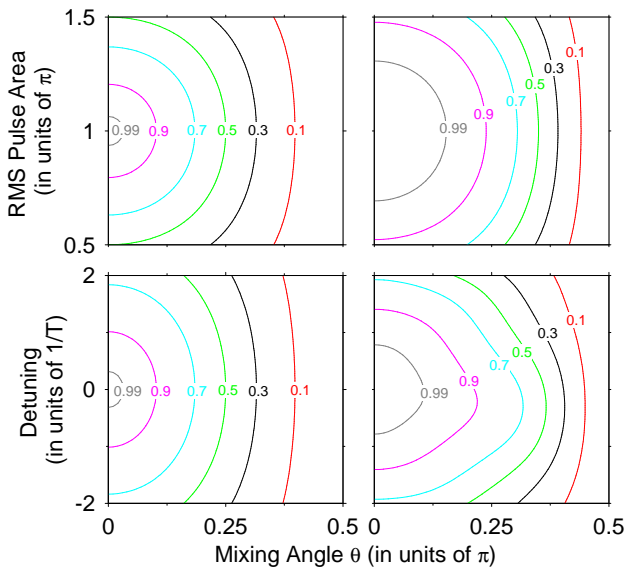}
 \end{center}
 \caption{(color online)
Transition probability $P_{1\to 2}$ for a V system.
\emph{Upper frames}: $P_{1\to 2}$ vs the mixing angle $\theta$ and the \rms~pulse area $A$
 for a single resonant pulse (upper left) and a composite sequence of three resonant pulses (upper right) with phases $\mathbf{\phi}_{12}=(0, 2/3, 0)\pi$; $\mathbf{\phi}_{13}=(0, 1, 1/3)\pi$.
\emph{Lower frames}: $P_{1\to 2}$ vs the mixing angle $\theta$ and the single-photon detuning $\Delta$ for a single rectangular pulse of duration $T$ and \rms~area $A=\pi$ (lower left)
 and a composite sequence of three rectangular pulses, each with duration $T$ and \rms~area $\pi$, and phases $\mathbf{\phi}_{12}=(0, 1/3, 0)\pi$; $\mathbf{\phi}_{13}=(0, 2/3, 0)\pi$ (lower right).
}
 \label{Cplot_all}
\end{figure}

Next, we have designed composite sequences which compensate \emph{simultaneous} deviations in the \rms~pulse area $A$ from $\pi$ and the mixing angle $\theta$ from 0.
In order to find the composite phases we use the Taylor expansion of $P_{1\to 2}$ with respect to both $A$ and $\theta$,
 and annul the coefficients of as many successive terms as possible, while requiring also $P_{1\to 2}= 1$ for $A=\pi$ and $\theta=0$.
This double compensation vs $A$ and $\theta$ is shown in Fig.~\ref{Cplot_all} (top frames) for a sequence of three resonant pulses.

In a similar manner, we have designed composite pulses which compensate simultaneous deviations in the mixing angle $\theta$ from 0 and the single-photon detuning $\Delta$ from resonance.
This double compensation is demonstrated in Fig.~\ref{Cplot_all} (bottom frames) for constant detuning: even a three-pulse composite sequence greatly expands the high-fidelity area around the desired point $(\theta=0,\Delta=0)$.

We note that even a triple compensation --- vs $\theta$, $A$ and $\Delta$ --- can be performed in a similar manner as above; it is, however, more difficult to illustrate it.

\sec{Y system.}
The method for suppression of unwanted transition paths is readily extended to more complex systems.
Here we describe a natural extension of the V system to a Y-shaped system in which a state $\ket{0}$ linked to state $\ket{1}$ is added, as shown in Fig.~\ref{Fig-VY} (right).
The RWA Hamiltonian reads
\begin{align}
\mathbf{H}_Y(t) &= \mathbf{H}_V(t) - (\hbar/2) \Delta \proj_{00}  \notag\\
 & + (\hbar/2)\left[\Omega_{01}(t)\e^{\i\phaseV_{01}} \proj_{01} + \text{h.c.} \right],
  \label{HamiltonianY1}
\end{align}
with $\mathbf{H}_V(t)$ being the Hamiltonian \eqref{HamiltonianV1} of the V system.
The coupling of the additional transition $\ket{0}\leftrightarrow\ket{1}$ is parameterized by a Rabi frequency with magnitude $\Omega_{01}(t)$,
 which should share the same time dependence $f(t)$ as the other two Rabi frequencies,
 and phase $\phaseV_{01}$ (which provides an additional control parameter).
In addition to the mixing angle $\theta$ in the V system,
 we introduce a second mixing angle $\xi$: $\Omega_{01}(t) = \Omega \sin\xi f(t)$, $\Omega_{12}(t) = \Omega \cos{\xi} \cos{\theta} f(t)$, and $\Omega_{13}(t) = \Omega \cos{\xi} \sin{\theta} f(t)$,
 where now $\Omega f(t)=\sqrt{\Omega_{01}(t)^2+\Omega_{12}(t)^2+\Omega_{13}(t)^2}$.
Hereafter we take $\xi=\pi/4$, i.e., $\Omega_{01}(t)^2 = \Omega_{12}(t)^2+\Omega_{13}(t)^2$.
The couplings in the Y system in each interaction step can be caused by the simultaneous application of two lasers, one on the lower transition $\ket{0}\to\ket{1}$ and another (elliptically polarized) on the upper V system $\ket{3}\from\ket{1}\to\ket{2}$. Alternatively, if in Fig.~\ref{Fig-VY} we bring state $\ket{0}$ up to the set of states $\ket{2}$ and $\ket{3}$, the Y system can be viewed as a tripod system; if the latter is formed of the magnetic sublevels in the transition $j=0\fromto j=1$ then the three couplings can be produced by a pair of pulses, with elliptical and linear polarizations, derived from the same laser.

The objective now is to transfer the population from state $\ket{0}$ to state $\ket{2}$ along the path $\ket{0}\to\ket{1}\to\ket{2}$, while suppressing the transition path $\ket{1}\to\ket{3}$.
Mathematically, this requires $|U_{20}|^2 = 1$,  $U_{00}=U_{10}=U_{30}=0$.

As in the V system, because the couplings share the same time dependence $f(t)$ and the Y system is on two-photon resonance, it can be transformed by a MS transformation into a set of two decoupled states and a two-state system.
This allows us to obtain an exact analytic expression for the propagator in the original basis in terms of the Cayley-Klein parameters of the MS two-state system, similar to the one of Eq.~\eqref{HamiltonianVMS} \cite{MorrisShore,Vitanov98,KyosevaVitanov,Genov11PRA}.
The propagator $\U$ for the $k$-th pulse pair depends now on three phases: $\phase_{01}^{(k)}$, $\phase_{12}^{(k)}$ and $\phase_{13}^{(k)}$.

\begin{figure}[tb]
\begin{center}
 \includegraphics[width=0.9 \columnwidth]{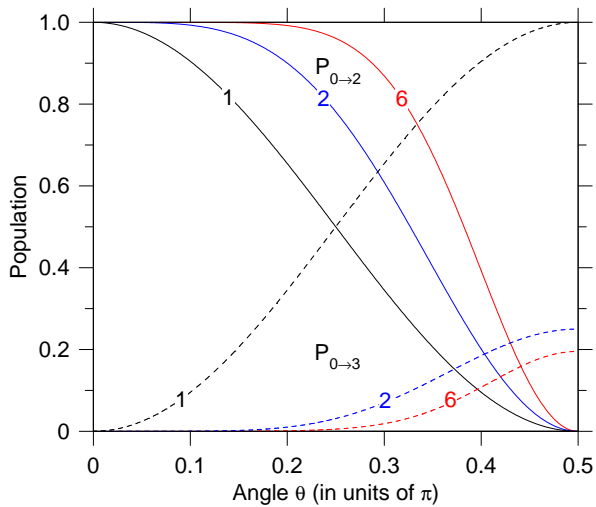}
 \end{center}
 \caption{(color online) Transition probabilities $P_{0\to 2}$ (solid curves) and $P_{0\to 3}$ (dashed curves) in a Y system vs the mixing angle $\theta$
  for a single pulse pair with \rms~area $2\pi$,
  and for composite sequences of two and six pulse pairs (each with \rms~area $\pi$), with phases:
    $\mathbf{\phi}_{01}=\mathbf{\phi}_{12}=(0,0)$ and $\mathbf{\phi}_{13}=(0,1)\pi$ for two pairs, and
    $\mathbf{\phi}_{01} = (0, 0, -0.181, -0.181, -0.033, -0.033)\pi$, $\mathbf{\phi}_{12} = (0, 0, -0.517, -0.517, -0.398, -0.398)\pi$, and $\mathbf{\phi}_{13} = (0, 0.562, 0.026, -1.554, 0.393, 0.238)\pi$ for six pairs.
  }
 \label{Fig-Y}
\end{figure}

Several conditions must be satisfied in order to achieve the desired transfer $\ket{0}\to\ket{2}$.
When the mixing angle is $\theta=0$ (no unwanted coupling $\ket{1}\to\ket{3}$),
 this is achieved by a pair of simultaneous resonant pulses with \rms~area $A=\int_{t_i}^{t_f} \Omega f(t) \d t=2\pi$ \cite{Vitanov98}. As in the V system, nonzero $\theta$ can be compensated, even without knowing its value, by a composite sequence of pulses.

The propagator of a sequence of $n$ pulse pairs is given by Eq.~\eqref{U_Comp} where now $\bm{\phi}_{k}=(\phi_{01}^{(k)},\phi_{12}^{(k)},\phi_{13}^{(k)})$. Composite sequences, which compensate deviations of $\theta$ and the other interaction parameters, are constructed in the same manner as for the V system above: we expand the transition probability $P_{0\to2}=|U_{20}^{(n)}|^2$ in a Taylor series vs the relevant parameter(s) and determine the composite phases from the condition to annul as many terms (in ascending order) in this expansion as possible.

\begin{figure}[tb]
\begin{center}
 \includegraphics[width=0.95\columnwidth]{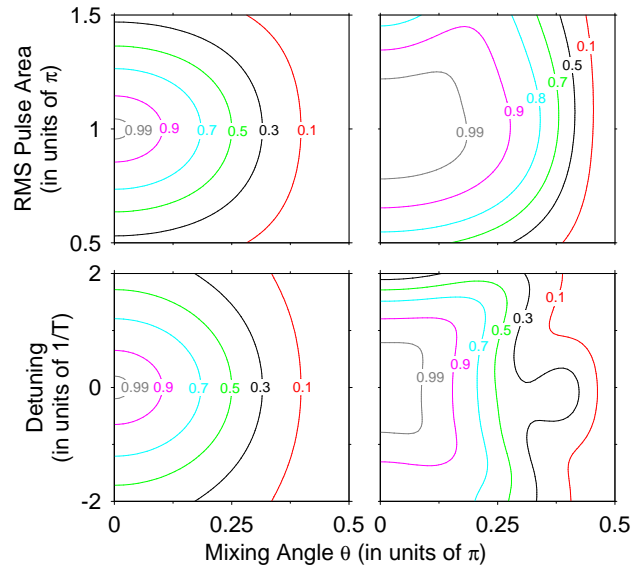}
 \end{center}
 \caption{(color online)
Transition probability $P_{0\to 2}$ for a Y system. \emph{Upper frames}: $P_{0\to 2}$ vs the mixing angle $\theta$ and the \rms~pulse area $A$
 for a single resonant pulse pair (upper left) and a composite sequence of six resonant pulse pairs (upper right) with phases $\mathbf{\phi}_{01}=(0, 0, -0.986, -0.986, 0.348, 0.348)\pi$, $\mathbf{\phi}_{12}=(0, 0, 0.667, 0.667, -0.317, -0.317)\pi$, and $\mathbf{\phi}_{13}=(0, -0.661, 0.337, -0.042, 0.955, 0.285)\pi$.
\emph{Lower frames}: $P_{0\to 2}$ vs the mixing angle $\theta$ and the constant single-photon detuning $\Delta$ for a single pair of rectangular pulses with \rms~area $A=\pi$ (lower left)
 and a composite sequence of 6 pairs of rectangular pulses, each with \rms~area $\pi$, and phases $\mathbf{\phi}_{01}=(0, 0, 0, 0, 4/3, 4/3)\pi$, $\mathbf{\phi}_{12}=(0, 0, 2/3, 2/3, 2/3, 2/3)\pi$, and $\mathbf{\phi}_{13}=(0, 0.937, 0.854, 0.171, 0.798, 1.448)\pi$.
}
 \label{Fig-YplotA}
\end{figure}

Compensation vs the mixing angle $\theta$ is demonstrated in Fig.~\ref{Fig-Y} for composite sequences of 2 and 6 pulse pairs. Simultaneous compensation of deviations in both the mixing angle $\theta$ from 0 and the \rms~pulse area $A$ from $\pi$
 is demonstrated in Fig.~\ref{Fig-YplotA} (upper frames) for a composite sequence of 6 pulse pairs.
Simultaneous compensation of deviations in the mixing angle $\theta$ from 0 and the single-photon detuning $\Delta$ from 0
 is demonstrated in Fig.~\ref{Fig-YplotA} (lower frames) for a composite sequence of 6 pulse pairs.

\sec{Conclusion.} \label{Concl}
The proposed technique is a simple and efficient method for robust population transfer and suppression of unwanted transition channels in multistate quantum systems.
We have demonstrated this technique for three-state V and four-state Y systems, but it can readily be adapted to more complex systems.
Unwanted transition channels may be merely unavoidable (e.g. due to off-resonant couplings), or can be activated, for instance, by deviations in light polarization or stray fields.
By suitably choosing the phases of the constituent pulses, the unwanted transition channels are suppressed with very high fidelity even when the corresponding couplings are \emph{unknown},
 because the technique suppresses the unwanted transitions for a broad range of coupling values.
Compensation for deviations in the experimentally controllable parameters --- polarizations, laser frequencies and laser intensities --- can be done simultaneously with respect to several of them.
The accuracy and the efficiency of the proposed technique and its experimental feasibility make it a potentially important tool in applications requiring high fidelity of operation, such as quantum information processing and quantum optics.

\acknowledgments
This work is supported by the European ITN project FASTQUAST, 
 and the Bulgarian NSF grants D002-90/08 and DMU-03/103.
We thank Daniel Comparat for stimulating discussions.



\begin{thebibliography}{99}

\bibitem{Levitt86} M. H. Levitt, Prog. NMR Spectrosc. \textbf{18}, 61 (1986).

\bibitem{Freeman97} R. Freeman, \emph{Spin Choreography} (Spektrum, Oxford, 1997).

\bibitem{Haffner} H.~H\"{a}ffner, C.F.~Roos, and R.~Blatt, Phys. Rep. \textbf{469}, 155 (2008).

\bibitem{RoosMolmer} I.~Roos and K.~M\o lmer, Phys. Rev. A, \textbf{69}, 022321 (2004).

\bibitem{Hill} C.D.~Hill, Phys. Rev. Lett., \textbf{98}, 180501 (2007).

\bibitem{Ivanov11PRA} S. S. Ivanov and N. V. Vitanov, Phys. Rev. A \textbf{84}, 022319 (2011).

\bibitem{Torosov11PRA} B.T.~Torosov and N.V.~Vitanov, Phys. Rev. A 83, 053420(7) (2011)

\bibitem{Ivanov11OL} S. S. Ivanov and N. V. Vitanov, Opt. Lett. \textbf{36}, 7 (2011).

\bibitem{Torosov11PRL} B.T.~Torosov, S. Gu\'erin and N.V.~Vitanov, Phys. Rev. Lett. \textbf{106}, 233001 (2011).

\bibitem{Genov11PRA} G.T.~Genov, B.T.~Torosov, and N.V.~Vitanov, Phys. Rev. A 84, 063413(10) (2011)

\bibitem{Optics} C. D. West and A. S. Makas, 
 J. Opt. Soc. Am. \textbf{39}, 791 (1949);
M. G. Destriau and J. Prouteau, 
 J. Phys. Radium \textbf{10}, 53 (1949);
S. Pancharatnam, 
 Proc. Ind. Acad. Sci. \textbf{51}, 130 (1955);
 \emph{ibid.} \textbf{51}, 137 (1955);
S. E. Harris, E. O. Ammann, and A. C. Chang, 
 J. Opt. Soc. Am \textbf{54}, 1267 (1964);
C. M. McIntyre and S. E. Harris, 
 \textit{ibid.} \textbf{58}, 1575 (1968).

\bibitem{WolfAzzamGoldstein}
 M. Born and E. Wolf, \emph{Principles of Optics} (Pergamon, Oxford, 1975);
 M. A. Azzam and N. M. Bashara, \emph{Ellipsometry and Polarized Light} (North Holland, Amsterdam, 1977);
 D. Goldstein and E. Collett, \emph{Polarized Light} (CRC Press, 2003).

\bibitem{Steffen03} M. Steffen, J. M. Martinis, and I. L. Chuang, Phys. Rev. B \textbf{68}, 224518 (2003)

\bibitem{WolfKlein} M. Born and E. Wolf, \emph{Principles of Optics} (Pergamon, Oxford, 1975); M. V. Klein, \emph{Optics} (John Wiley \& Sons, New York, 1970).

\bibitem{MorrisShore} J.R.~Morris and B.W.~Shore, Phys. Rev. A \textbf{27}, 906 (1983).

\bibitem{Vitanov98} N. V. Vitanov, J. Phys. B  \textbf{31}, 709 (1998).

\bibitem{KyosevaVitanov} E.S.~Kyoseva and N.V.~Vitanov, Phys. Rev. A \textbf{73}, 023420 (2006).

\end{thebibliography}
\end{document}